\documentclass[conference]{IEEEtran}
\IEEEoverridecommandlockouts
\usepackage{enumitem}
\usepackage{cite}
\usepackage{amsmath,amssymb,amsfonts}
\usepackage{algorithmic}
\usepackage{enumitem}
\usepackage{graphicx}
\usepackage{textcomp}
\usepackage{xcolor}
\usepackage{csquotes}
\usepackage{balance}
\def\BibTeX{{\rm B\kern-.05em{\sc i\kern-.025em b}\kern-.08em
    T\kern-.1667em\lower.7ex\hbox{E}\kern-.125emX}}

\usepackage[acronym,toc,shortcuts]{glossaries}

\graphicspath{{figures/}}

\makeglossaries
\newacronym{3GPP}{3GPP}{The 3rd Generation Partnership Project }
\newacronym{5G}{5G}{Fifth Generation}
\newacronym{AAA}{AAA}{Authentication, Authorization and Accounting}
\newacronym{ARIMA}{ARIMA}{AutoRegressive Integrated Moving Average}
\newacronym{CPU}{CPU}{Central Processing Unit}
\newacronym{GRU}{GRU}{Gated Recurrent Unit}
\newacronym{O-RAN}{O-RAN}{Open Radio Access Network}
\newacronym{QoS}{QoS}{Quality of Service}
\newacronym{PRB}{PRB}{Physical Resource Block}
\newacronym{rApp}{rApp}{radio App}
\newacronym{RAN}{RAN}{Radio Access Network}
\newacronym{RIC}{RIC}{RAN Intelligent Controller}
\newacronym{xApp}{xApp}{eXtended application}
\newacronym{SFF}{SFF}{Simple-Feed-Forward}
\newacronym{LSTM}{LSTM}{Long-Short Term Memory}
\newacronym{SN}{SN}{Seasonal-Naive}
\newacronym{MLP}{MLP}{ Multi-Layer Perceptron}
\newacronym{RNN}{RNN}{Recurrent Neural Network}
\newacronym{MSE}{MSE}{Mean Square Error}
\newacronym{MASE}{MASE}{Mean Absolute Scaled Error}
\newacronym{MAPE}{MAPE}{Mean Absolute Percentage Error}
\newacronym{ND}{ND}{Normalized Deviation}
\newacronym{QL}{QL}{Quantile Loss}
\newacronym{AI}{AI}{Artificial Intelligence}
\newacronym{ML}{ML}{Machine Learning}
\newacronym{SLAs}{SLAs}{Service Level Agreements}

\usepackage{fancyhdr}
\usepackage{geometry}
\usepackage{lipsum} 

\begin{document}

\title{On the use of Probabilistic Forecasting for Network Analysis in Open RAN \\
}


\author{Vaishnavi Kasuluru, Luis Blanco, Engin Zeydan \\
{\normalsize{} Centre Technologic de Telecomunicacions de Catalunya (CTTC/CERCA), Castelldefels, Barcelona, Spain, 08860.} \\
{\normalsize{} Emails: \texttt{\{vkasuluru, lblanco, ezeydan\}@cttc.es}}

\thanks{This work has been supported by SEMANTIC project, funded by the European Union’s Horizon 2020 research and innovation program under the Marie Skłodowska-Curie grant (agreement No 861165), the Horizon Europe project VERGE (ID: 101096034), the Spanish projects FREE6G-RadEdge (TSI-063000-2021-121) and FREE6G-RegEdge (TSI-063000-2021-144) funded by MINECO through the “NextGenerationEU” program, and the Spanish project ORIGIN (PID2020-113832RB-C22) funded MICCIN.}

}

\maketitle

\begin{abstract}

Unlike other single-point Artificial Intelligence (AI)-based prediction techniques, such as Long-Short Term Memory (LSTM), probabilistic forecasting techniques (e.g., DeepAR and Transformer) provide a range of possible outcomes and associated probabilities that enable decision makers to make more informed and robust decisions. At the same time, the architecture of Open RAN has emerged as a revolutionary approach for mobile networks, aiming at openness, interoperability and innovation in the ecosystem of RAN. In this paper, we propose the use of probabilistic forecasting techniques as a radio App (rApp) within the Open RAN architecture. We investigate and compare different probabilistic and single-point forecasting methods and algorithms to estimate the utilization and resource  demands of Physical Resource Blocks (PRBs) of cellular base stations. Through our evaluations, we demonstrate the numerical advantages of probabilistic forecasting techniques over traditional single-point forecasting methods and show that they are capable of providing more accurate and reliable estimates. In particular, DeepAR clearly outperforms single-point forecasting techniques such as LSTM and Seasonal-Naive (SN) baselines and other probabilistic forecasting techniques such as Simple-Feed-Forward (SFF) and Transformer neural networks.




\end{abstract}


\section{Introduction}
\label{intro}

\ac{O-RAN} is a new communications paradigm that is expected to provide the technological foundation for the next generation of communications systems. \ac{O-RAN} architecture has emerged as a transformative approach to mobile networks that aims to introduce openness, interoperability and innovation within the RAN ecosystem\cite{polese2023understanding}. \ac{O-RAN} promotes the disaggregation of network components and the use of open interfaces, enabling greater vendor diversity and flexibility in network deployment. In the context of Open RAN, the concept of \glspl{rApp} plays a critical role in enhancing the capabilities of \ac{RIC} and enabling advanced functionalities. In B5G and 6G systems, \ac{O-RAN} offers a plethora of important benefits, such as high spectral efficiency, flexibility, uniform \ac{QoS} in a given area, an open ecosystem, and a vendor-neutral network that readily enables the aggregation of smart, third-party services.




This paper addresses the integration of probabilistic forecasting as \ac{rApp} within the architecture of \ac{O-RAN}. In the context of \ac{RAN}, network and computational resource prediction enables better baseband processing, radio resource management, mobility management, beamforming, scheduling, and other tasks required for \ac{RAN} operation. Probabilistic forecast refers to the use of probabilistic methods and algorithms to estimate the expected resource demands (e.g., \ac{CPU} or \ac{PRB} utilization) of various network functions and services within the \ac{RAN}. By leveraging historical data, statistical models, and advanced algorithms, probabilistic forecast provides valuable insight into resource allocation and sharing among entities, enabling efficient use of computing resources while meeting \ac{QoS} requirements. Therefore, the use of probabilistic forecast (e.g., \ac{CPU} or \ac{PRB}) in \ac{RAN} can enable better capacity planning and resource management, allowing operators to anticipate and address potential resource constraints or over-provisioning issues.

\subsection{Related Work}

The upcoming fifth (5G) and sixth (6G) iterations of mobile networks will speed up the transformation from rigid and consolidated networks to dynamic, decentralized structures. These advanced architectures will rely on the implementation of software-based solutions, virtualization, as well as the principles of openness and the ability to reprogram network components \cite{polese2023understanding}. To ensure compatibility and facilitate the utilization and advancement of management and operation software within a virtualized \glspl{RAN}, the industry is embracing the concept of Open RAN \cite{kliks2023towards}. \ac{O-RAN} architecture is specifically designed to be adaptable, scalable, and driven by software, making it easier to deploy, maintain, and optimize \glspl{RAN} networks \cite{lacava2023programmable}. Through the implementation of \glspl{RAN} virtualization, Mobile network operators can effectively share \glspl{RAN} resources, resulting in cost reduction and improved energy efficiency. Several studies and research papers have explored various aspects of \ac{O-RAN}, addressing different challenges and proposing innovative solutions \cite{hammami2022policy}. These related works cover a wide range of topics, including resource allocation, network optimization, and intelligent algorithms. For example, some papers focus on using \ac{AI} methods to tackle the challenge of radio and computing resource allocation to minimize overall network latency\cite{hammami2022policy}, monitoring the \ac{SLAs} of each network slice and forecasting its future resource usage in the \glspl{RAN} \cite{gutterman2019ran} and assessing different statistical metrics derived from the utilization data of \ac{PRB} to explore their predictability in order to develop deep learning models that can forecast the utilization of \ac{PRB} in \glspl{RAN} \cite{nagib2021deep}.

\par

Effective resource utilization in \glspl{RAN} is critical to achieving multiplexing gains and cost efficiencies for service providers. Overestimation of resource allocation leads to lower revenues for the service provider, while underestimation leads to violations of \ac{SLAs} \cite{gutterman2019ran}. The number of solutions that use deterministic single point forecasting models to predict network conditions and radio resource requirements has been growing\cite{10071958}. The majority of studies use either statistical methods such as \ac{ARIMA} or \ac{ML} methods such as \ac{LSTM} neural networks. By accounting for uncertainty and variability in the data, probabilistic forecasting can provide more accurate predictions compared to deterministic forecasting methods. It provides a range of possible outcomes along with their associated probabilities, allowing for a more realistic representation of future events. By understanding the probabilities associated with different scenarios, they can allocate resources, optimize network performance, and deliver services more effectively while meeting user demands \cite{li2017intelligent}. Probabilistic forecasting offers valuable insights and benefits to service providers. However, it is important to carefully consider data requirements, complexity, computational resources, and how to deal with external uncertainties to effectively leverage the benefits and mitigate potential limitations. It is also critical to develop an accurate forecasting model to ensure optimal use of \glspl{RAN}.





\subsection{Contributions}


This paper is primarily concerned with evaluating a probabilistic forecasting approach as an \ac{rApp} within the \ac{O-RAN} architecture. Specifically, we aim to investigate and compare different probabilistic methods and algorithms to estimate the expected utilization or resource demands of \glspl{PRB} for different network functions and services within the \ac{O-RAN} architecture. By leveraging probabilistic forecasting techniques, we show how mobile operators can gain valuable insights to make informed resource allocation decisions in the specific case of \ac{PRB} sharing. Through our evaluations, we demonstrate the numerical advantages of using probabilistic forecasting techniques over traditional methods and show that they can provide more accurate and reliable estimates of \ac{PRB} utilization. In particular, DeepAR achieves the lowest \ac{MSE} value of 0.065, the lowest \ac{MASE} value of 0.175, and a significantly lower \ac{MAPE} value of 0.016 compared to probabilistic techniques such as \ac{SFF}, Transformer, \ac{LSTM}, and \ac{SN} baselines. To the best of our knowledge, this is the first work to investigate the applicability of probabilistic forecasting  to be offered as \ac{rApp} in the Open RAN architecture.




\section{Background Information}

\subsection{\ac{O-RAN} Architecture and Components}


The dependence of traditional \ac{RAN} on hardware components and the vendor lock-in is causing huge CAPEX and OPEX costs, which in turn hinder the creation of an intelligent, collaborative, and reliable network. Therefore, it is important to develop the next generation of \ac{RAN} solutions based on universal, vendor-independent hardware and software-defined technology. Virtualization and \ac{RAN} dis-aggregation are the key technologies for the concept of \ac{O-RAN}, whose main pillars are openness and intelligent resource management. The main components of \ac{O-RAN} are:

\begin{itemize}
    \item Open-Radio Unit (O-RU), Open-Distributed Unit (O-DU) and Open-Central Unit (O-CU) whose functionalities are similar to that in 5G dis-aggregated RAN except with added support of \ac{O-RAN} based specifications and interface.
    \item Near-Real Time \ac{RIC}  for control/optimization of RAN elements and resources based on fine grained data using online \ac{AI}/\ac{ML}. It is suitable for application with latency needs of between 10ms and 1s.
    \item Non-Real Time \ac{RIC} for Control/optimization of RAN elements and resources based on coarse grained data using online \ac{AI}/\ac{ML}. It is suitable for application with latency requirement greater than 1s. It also provides policy based guidance to near-Real Time \ac{RIC}.
    \item \ac{RIC} components are deployed at some point between the cell site and the core network. They can be placed either in the edge or in the regional cloud network, depending upon the usage scenario.
\end{itemize}

\begin{figure}[htp!]
\centering
\includegraphics[width=\linewidth]{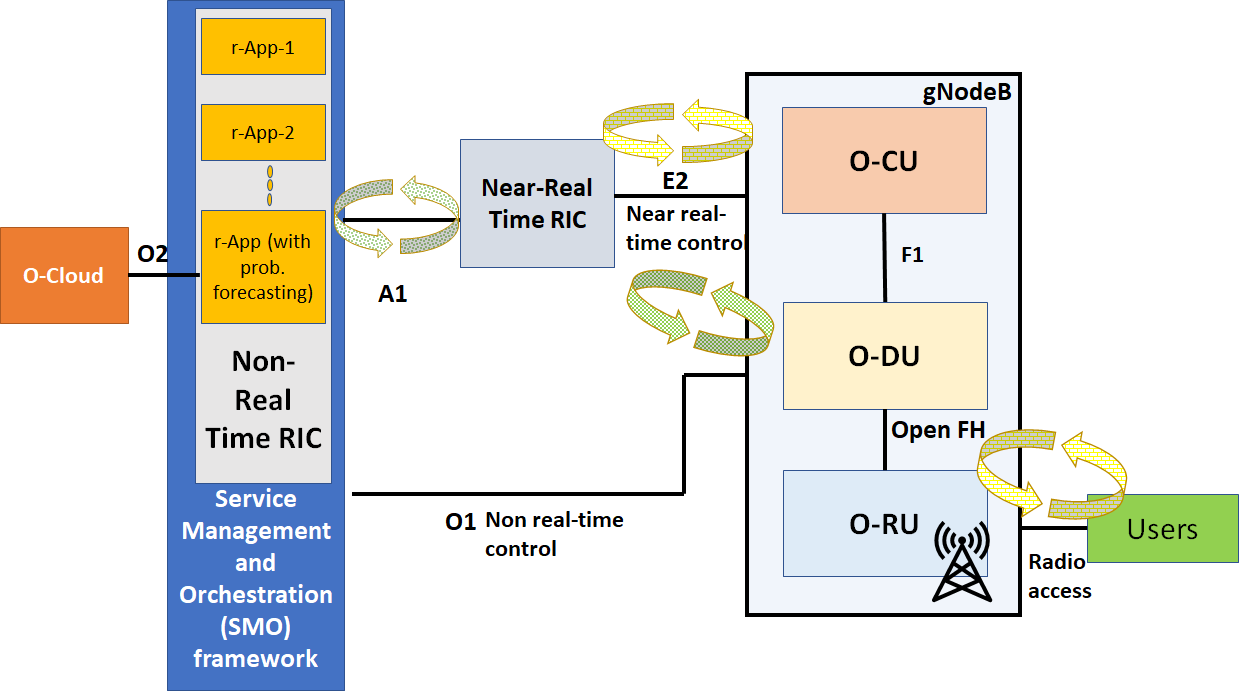}
\caption{Open RAN architecture with probabilistic forecast as rApp}
\label{rapp}
\end{figure}

Fig. \ref{rapp} shows the open \ac{RAN} architecture with probabilistic forecast as \ac{rApp}. Probabilistic forecasting can play a valuable role in the \ac{PRB} allocation process within the \ac{rApp} framework. Using historical data, statistical models, and advanced algorithms, probabilistic forecasting can provide insight into the expected resource requirements of various applications and services over time.By integrating probabilistic forecasting into the \ac{PRB} allocation process as an \ac{rApp}, operators can optimize resource utilization, improve network efficiency, and increase the overall performance of the radio access network. The dynamic nature of probabilistic forecasting enables operators to adapt and respond to changing network conditions and resource requirements in real time, ensuring effective \ac{PRB} allocation for various applications and services.

\subsection{Probabilistic forecast and motivation}

Probabilistic time series forecasting is an important tool that provides a range of possible outcomes and their associated probabilities instead of a single point estimate. There has also been much recent innovation in forecasting methods, with models based on deep learning being used alongside classical methods. An important aspect that highlights the advantages of probabilistic forecasting techniques over classical \ac{AI}-based algorithms (e.g., \glspl{LSTM}, \glspl{GRU}, etc.) or classical techniques (e.g., \ac{ARIMA}, exponential smoothing) is the problem of overconfidence in pointwise prediction values. Classical data-driven, \ac{AI}-based models often provide point forecasts, giving a false sense of confidence in the accuracy of their predictions. This overconfidence arises because these models fail to capture the uncertainties and errors associated with their forecasts. The limitations of point forecasts can create a significant information gap in the decision-making process. When decision- makers or decision engines rely solely on point forecasts without considering the uncertainties associated with them, they may make decisions that are inadequate or misguided.This information gap leads to a range of possibilities, and the magnitude of this range increases as the information gap increases. Probabilistic forecasting is proposed to solve this problem. Unlike point forecasts, probabilistic forecasts produce a predictive distribution of values rather than a single point estimate. This distribution represents the range of possible outcomes along with their respective probabilities. By providing a comprehensive view of the uncertainties and capturing the full range of possibilities, probabilistic forecasting enables decision makers to make more informed and robust decisions. Probabilistic forecasting techniques offer several reasons for their use over classical AI-based forecasting algorithms such as \glspl{LSTM} or \glspl{GRU}. Some reasons why probabilistic forecasting is better than classical time series forecasting:



\textbf{(i) Uncertainty quantification:} Probabilistic forecasting provides a way to quantify uncertainty in predictions by providing a range of possible outcomes and their associated probabilities. Unlike deterministic forecasts, which provide a single point estimate, probabilistic forecasting techniques generate a range of possible outcomes along with their associated probabilities. 



\textbf{(ii) Decision support:} The probabilistic nature of forecasting allows for better decision support. Decision-makers can assess the potential risks and opportunities associated with different outcomes based on the probability distribution provided by the probabilistic forecast. This enables more informed decision-making, especially in situations where the consequences of different outcomes vary significantly.




\textbf{(iii) Model interpretability:} Probabilistic forecasting techniques often provide more interpretability compared to black-box models like \ac{LSTM}s or GRUs. The explicit modeling of uncertainty allows for a clearer understanding of how different factors contribute to the forecast. 

\textbf{(iv) Model robustness:} Probabilistic forecasting techniques offer greater robustness against outliers and anomalies in the data. Classical AI-based algorithms can be sensitive to extreme values, leading to skewed or biased forecasts. In contrast, probabilistic models, by considering the entire distribution of possible outcomes, are more resilient to the impact of outliers and can provide more accurate forecasts even in the presence of anomalous data points.
 
\textbf{(v) Handling non-stationarity:} Traditional AI-based forecasting algorithms like \ac{LSTM}s or GRUs assume stationarity in the underlying data, meaning that the statistical properties of the data remain constant over time (i.e., trend in the data is not mean-reverting). However, real-world data often exhibit non-stationary patterns, where the statistical properties change over time. Probabilistic forecasting techniques can better handle non-stationarity by explicitly modeling and updating the uncertainty as new data becomes available.
 


In summary, probabilistic forecasting techniques offer valuable advantages over classical AI-based forecasting algorithms such as \ac{LSTM}s or GRUs, including uncertainty quantification, decision support, handling non-stationarity, model robustness, better  uncertainty management and mitigation of risks. These motivations make probabilistic forecasting techniques a compelling choice in many real-world forecasting scenarios.

\section{Review of probabilistic Forecasting}

\subsection{Forecasting Techniques}

Probabilistic forecasting techniques, such as \textbf{\ac{SFF}, DeepAR and Transformers}, are becoming increasingly popular due to their ability to provide more accurate and reliable forecasts by taking into account uncertainty and providing a probabilistic distribution of future values. \textbf{\ac{SFF}} estimator refers to a type of predictive model that uses a feed-forward neural network to estimate probabilities for different outcomes. A feed-forward neural network, also known as a \ac{MLP}, consists of an input layer, one or more hidden layers, and an output layer. Each layer is composed of multiple neurons or units that perform computations. The connections between neurons are directed and propagate forward, from the input layer to the output layer, without any loops or cycles. During training, the feed-forward network learns to map the input features to the corresponding probabilities. This learning process involves adjusting the weights and biases of the neurons based on the error between the predicted probabilities and the true probabilities from the training data. This adjustment is typically done using backpropagation, where the error is propagated backward through the network, and the gradients are computed to update the weights. Once trained, the \ac{SFF} estimator can take new input features and produce predicted probabilities for the different outcomes. These probabilities can be interpreted as the model's confidence in each outcome, allowing for uncertainty quantification in the predictions.

\textbf{DeepAR} is a probabilistic forecasting technique developed by Amazon that is based on a \ac{RNN} architecture. It is designed to forecast time series data by learning patterns and dependencies in the historical data, and then generating a probabilistic distribution of future values\cite{salinas2020deepar}. DeepAR works by taking in a time series as input and using an RNN to model the temporal dependencies between the observations. The \ac{RNN} uses a combination of past observations and covariates (such as weather or economic data) to predict future values. The output of the \ac{RNN} is then passed through a fully connected layer to generate a probabilistic distribution of future values. One of the strengths of DeepAR is its ability to model both seasonal and non-seasonal patterns in the data. It also allows for the incorporation of covariates, which can help to improve the accuracy of the forecasts. DeepAR uses a negative log-likelihood loss function to optimize the parameters of the neural network during training. This loss function takes into account both the mean and variance of the predicted distribution, and encourages the model to generate forecasts that are both accurate and uncertain.

\textbf{Transformers} are a type of neural network architecture that have been adapted for probabilistic forecasting of time series data. The Transformer architecture was originally developed for natural language processing tasks, but has since been applied to time series forecasting due to its ability to handle long-term dependencies and capture complex patterns in the data\cite{vaswani2017attention}. The Transformer architecture consists of a series of encoder and decoder layers, which use self-attention to identify relevant time steps and combine them in a flexible and adaptive way. The encoder layers process the input sequence, while the decoder layers generate the forecasted sequence. The self-attention mechanism allows the model to attend to different parts of the input sequence at different times, depending on the relevance to the current forecast. In order to generate a probabilistic forecast, the Transformer model produces a probability distribution over the possible future values, which can be sampled to produce a range of possible outcomes and their associated probabilities. 



Overall, \ac{SFF} estimators are simple to implement and understand, suitable for small to medium-sized datasets, and can model complex non-linear relationships, but it lacks the ability to capture sequential dependencies in the data, which can be important for time series forecasting tasks. The \ac{SFF} estimator is efficient when the dataset is relatively small, and the temporal dependency is not crucial for accurate forecasting.  DeepAR and Transformer are powerful and flexible probabilistic forecasting techniques that can be used in a variety of applications. Its ability to model complex patterns in the data and generate probabilistic forecasts make it a valuable tool for decision-making and resource allocation. The main advantages of such probabilistic forecasting methods over classic time series forecasting like \ac{LSTM} is the ability to provide more accurate and reliable forecasts, by capturing the uncertainty associated with the forecast. This is particularly important in applications where there is a high degree of uncertainty or variability in the data. In addition, probabilistic forecasting can provide a more complete picture of the range of possible outcomes, which can be useful for decision-making and risk management.

\subsection{Evaluation Metrics}
Evaluation metrics for probabilistic time series forecasting typically focus on comparing the forecasted probability distribution $\hat{y_i}$ to the actual outcomes $y_i$. They help to evaluate the accuracy of the forecasts and identify areas for improvement. Some commonly used evaluation metrics for probabilistic time series forecasting:
\begin{itemize}[leftmargin=1.5mm]

\item \textbf{\ac{MSE}:} This measures the average of the squared differences between the forecasted and actual values. It is a commonly used metric for point forecasting, but can also be used for probabilistic forecasting by taking the mean of the squared differences between the forecasted and actual values over the forecasted distribution.
\begin{equation}
MSE = \frac{1}{N} \sum_{i=1}^{N} (Y_i-\hat{Y_i})^2
\end{equation}
\noindent where N is the number of forecasting points.

\item \textbf{\ac{MASE}:} This is a scaled version of the mean absolute error, which compares the forecasted values to a naive forecast (such as a historical average). It is often used to compare different models, as it scales the error by the error of the naive forecast.
\begin{equation}
MASE = \frac{1}{N} \sum_{i=1}^{N} |Y_i-\hat{Y_i}| 
\end{equation}

\item \textbf{\ac{MAPE}:} This measures the percentage difference between the forecasted and actual values. It is commonly used in business forecasting to evaluate the performance of models.
\begin{equation}
MAPE = \frac{1}{N} \sum_{i=1}^{N} (|Y_i-\hat{Y_i}| / {|Y_i|})
\end{equation}
\item \textbf{Coverage:} This measures the proportion of actual values that fall within the predicted interval. For example, a 90\% prediction interval should capture 90\% of the actual values. It is often used to evaluate the calibration of probabilistic forecasts.
\begin{equation}
 Coverage = \frac{1}{N} \sum_{i=1}^{N} (Y_i<\hat{Y_i}) 
\end{equation}

\item \textbf{\ac{ND}:} This measures the average deviation between the forecasted and actual values, normalized by the range of the actual values. It is useful for evaluating the accuracy of probabilistic forecasts, particularly when the range of the actual values is large.
\begin{equation}
ND = \frac{1}{N} \sum_{i=1}^{N} (|Y_i-\hat{Y_i}| / mean(Y_i))
\end{equation}

\item \textbf{\ac{QL}:} This measures the average difference between the predicted and actual quantiles of the forecasted distribution at given \ac{QL}. It is a commonly used metric for evaluating probabilistic forecasts, particularly in applications where accurate quantile estimates are important.


\begin{equation}
\text{QuantileLoss} = 
\begin{cases}
(Y_i-\hat{Y_i})*QL & \text{if } Y_i \geq \hat{Y_i} \\
(\hat{Y_i}-Y_i)*(1-QL) & \text{if } Y_i < \hat{Y_i}
\end{cases}
\end{equation}

\end{itemize}

\section{Simulation Results}


In this section, we examine the level of accuracy in predicting the outcomes for each suggested models and compare them to the deterministic estimators. The \ac{rApp} was created in Python using the Gluonts library\footnote{Online: https://ts.gluon.ai/stable/, Available: May 2023.}, which uses three different estimators (\ac{SFF}, DeepAR, and Transformer) for forecasting. It uses allocation data from \ac{PRB} which can be obtained through the O1 interface from O-CU in an \ac{O-RAN} architecture.The dataset was created by simulating traffic and mobility patterns for a varying number of end users\cite{9933014}. To evaluate the accuracy of our forecasts, we calculate metrics such as \ac{MSE}, \ac{MASE}, \ac{MAPE}, and \ac{ND} based on the true values. In addition, we evaluate the quality of probabilistic forecasts using \ac{QL}, which is calculated for forecast quantiles from 0.1 to 0.9. We also examine the calibration and sharpness of each model by measuring the coverage of true values within certain percentiles (e.g., from 0.1 to 0.9). A higher coverage level, closer to 0.8, is considered more desirable. These metrics, collectively referred to as Coverage[X] (e.g., Coverage[0.5]) when a specific coverage value of X is targeted, help us evaluate the accuracy of the models. Evaluating the values at the extreme percentiles (e.g., Coverage[0.9]) allows us to assess how well the models capture the tails of the distribution. To evaluate the accuracy of each estimator, the dataset is divided into training and test data. The training data is used to fit the machine learning parameters for each method, while the test data is used to evaluate and compare the performance of each model. 

In Table 1, we compare our probabilistic models with \ac{LSTM} and \ac{SN} baseline models. It is evident that the probabilistic models outperform \ac{LSTM} and \ac{SN} in terms of \ac{MSE}, \ac{MASE}, and \ac{MAPE}. This is because the deterministic baseline models are more susceptible to outliers, uncertainty, and nonstationary data, which can lead to biased or skewed forecasts. Among the models considered, DeepAR achieves the highest level of accuracy on all metrics (e.g., DeepAR achieves the lowest \ac{MSE} of 0.065, the lowest \ac{MASE} of 0.175, and a significantly lower \ac{MAPE} of 0.016). DeepAR also performs better in terms of \ac{ND},Coverage and \ac{QL} except for Coverage with 10\% prediction interval. This can be attributed to the fact that DeepAR's loss function takes into account both the mean and the variance of the predicted distribution, thus promoting the generation of forecasts that are both accurate and exhibit appropriate levels of uncertainty.


\begin{table}[htp!]
\vspace{-0.5cm}
\caption{Accuracy comparison }
\centering
\begin{tabular}{|c|c|c|c|c|c|}
\hline
\bf{ Metrics} & \bf{LSTM} & \bf{SN} & \bf{SFF} & \bf{DeepAR} & \bf{Transformer}\\
\hline
MSE  & 4.243 & 4.146  & 3.233  & \textbf{0.065}   & 0.906  \\
\hline
MASE  & 1.182   & 1.114  & 1.013    & \textbf{0.175}   & 0.624\\
\hline
MAPE  & 8.875   & 0.098  & 0.099 & \textbf{0.016}  & 0.058   \\
\hline
ND  & -   & 0.099  & 0.090   & \textbf{0.016}    & 0.055\\
\hline
QL[0.1]  & -   & 61.40  & 26.46   & \textbf{3.03}     & 20.47\\
\hline
Coverage[0.1]  & -   & 0.271  & 0.104   & 0.146     & \textbf{0.333}\\
\hline
QL[0.5]  & -   & 67  & 60.96   & \textbf{10.53}     & 37.55\\
\hline
Coverage[0.5]  & -   & 0.271 & 0.479   & \textbf{0.896}     & 0.625\\
\hline
QL[0.9]  & -   & 72.60  & 23.31   & \textbf{4.69 }    & 18.50\\
\hline
Coverage[0.9]  & -   & 0.271  & 0.938   & \textbf{1.00}     & 0.771\\
\hline
\end{tabular}
\end{table}

\begin{figure}[!h]
\centering
\includegraphics[width=\linewidth]{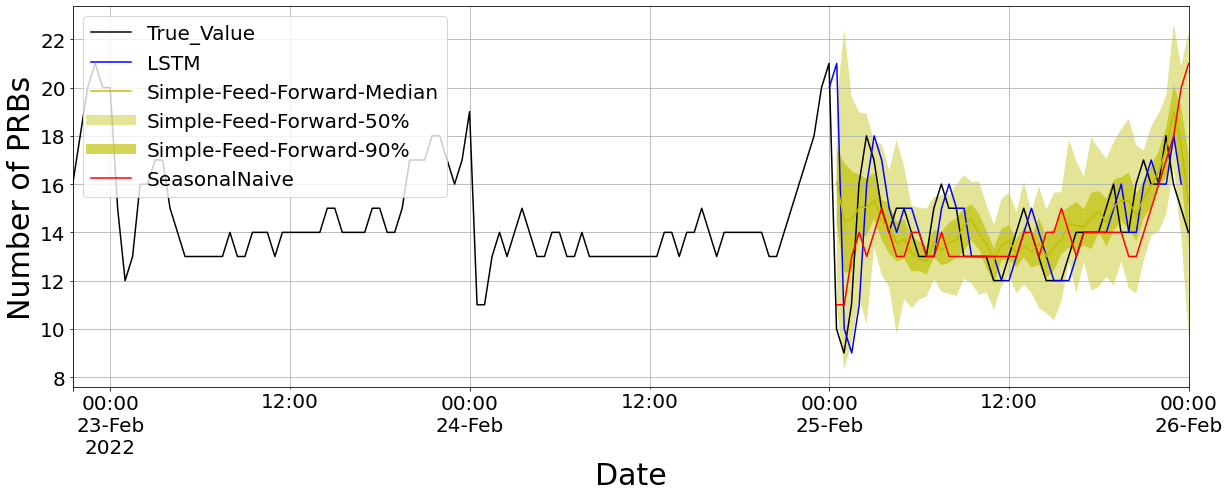}
\caption{Comparison of SFF estimator with deterministic single point estimators}
\label{sff_pred1}
\vspace{-0.5cm}
\end{figure}

\begin{figure}[!h]
\centering
\includegraphics[width=\linewidth]{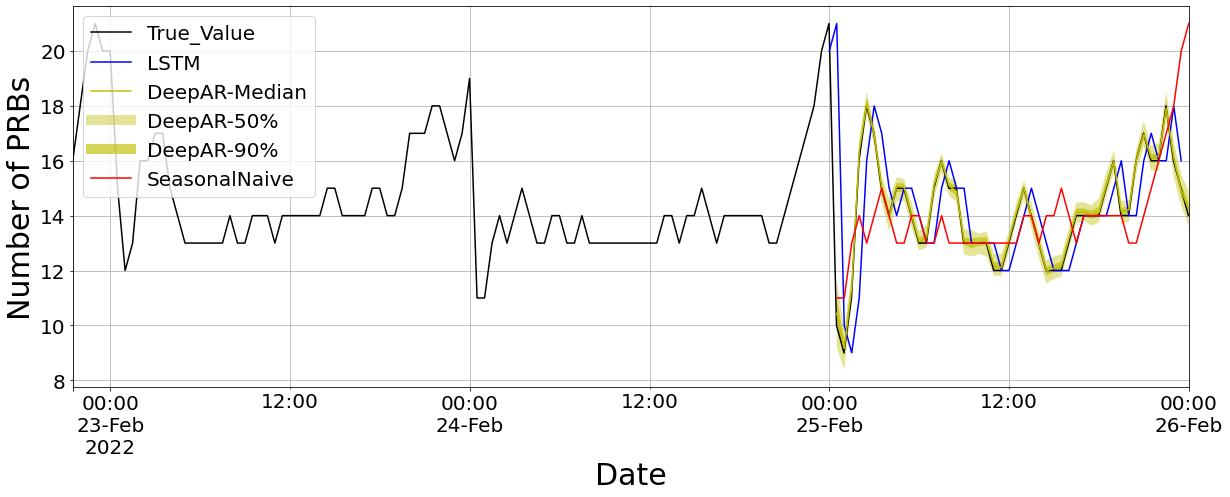}
\caption{Comparison of DeepAR estimator with  deterministic single point estimators}
\label{DA_pred1}
\end{figure}

\begin{figure}[!h]
\centering
\includegraphics[width=\linewidth]{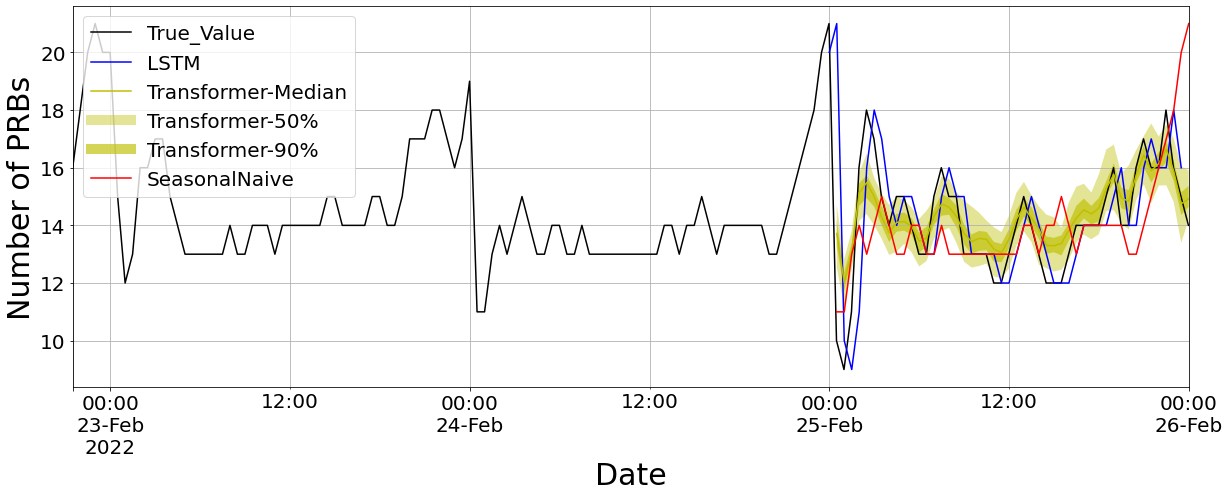}
\caption{Comparison of Transformer estimator with  deterministic single point estimators}
\label{T_Pred1}
\vspace{-0.5cm}
\end{figure}

Figures \ref{sff_pred1}, \ref{DA_pred1}, and \ref{T_Pred1} illustrate the comparison between probabilistic estimators and deterministic baseline estimators for forecasting future \ac{PRB} values. In these figures, true values are represented by the black line, \ac{LSTM} predictions by the blue line, SN predictions by the red line, and the olive green line represents the probabilistic estimators. From the figures, it can be seen that the predictions generated by the \ac{LSTM} and \ac{SN} models are the least accurate. Conversely, it can be observed that the predictions generated by the \ac{SFF} and DeepAR are superior to those of the Transformer model as well as the deterministic models. The combination of capturing uncertainty, accounting for temporal dependencies, and using similar time series for training makes DeepAR a preferred choice over the Transformer or other models for time series forecasting. However, the suitability of any model may also depend on the particular data set and problem at hand.

\begin{figure}[htp!]
\centering
\includegraphics[width=\linewidth]{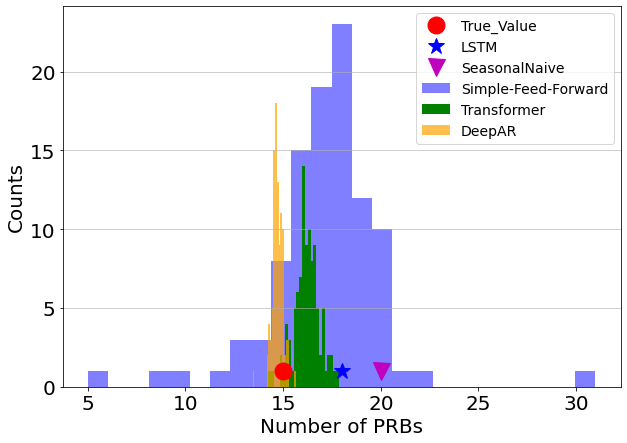}
\caption{Histogram of true value and estimators}
\vspace{-0.6cm}
\label{hist}
\end{figure}

Figure \ref{hist} shows the histogram of both the baseline and probabilistic estimators. It can be deduced that the predictions produced by the DeepAR and Transformer models are centered around the true value. However, the predictions of the \ac{SFF} model have a larger dispersion, which is due to a higher uncertainty in the forecasting process. The higher uncertainty in the predictions can be seen in Figure \ref{sff_pred1}. From Figure \ref{hist}, it can be seen that the predictions produced by the baseline models are significantly away from the true value.

\section{Main Observation and Discussion}
\label{observations}

The aim of this section is to briefly summarize the main observations and share lessons learnt from the training and the evaluation of probabilistic forecasting methods. Next, we provide some insights. One of the main observations is that when the dataset used for probabilistic forecast is very small, models like DeepAR may underfit. DeepAR is a deep learning-based method that learns to capture the temporal dependencies and generate probabilistic forecasts. However, with limited data, the model may not have enough information to capture the underlying patterns and generate accurate forecasts. Another important point is that training time for methods like DeepAR and Transformer can be high. These models often require significant computational resources and can take a long time to train, especially when dealing with large datasets. On the other hand, utilizing GPUs for training can significantly reduce the training time by leveraging their parallel processing capabilities.

The complexity of probabilistic forecast models can be a challenge. Neural networks used for probabilistic forecasting can have a high number of parameters, leading to increased computational complexity. To address this issue, techniques to reduce neural network complexity can be employed, such as model compression, pruning, or using lighter architectures without compromising the accuracy of the forecast. DeepAR, specifically, has been found to perform better with non-stationary datasets. Non-stationary datasets exhibit changing patterns and trends over time, and DeepAR's ability to capture temporal dependencies makes it well-suited for such scenarios. By considering the historical context, DeepAR can generate probabilistic forecasts that adapt to the changing dynamics of the dataset.


\section{Conclusions}

In this paper, we compared probabilistic forecasting techniques with single-point \ac{AI}-based prediction methods, such as. \ac{LSTM}, and investigated their applicability within the \ac{O-RAN} architecture as \ac{rApp}. Our results highlight the advantages of using probabilistic forecasting techniques, such as DeepAR and Transformer, over single-point forecasting methods in the context of \ac{PRB} forecasting in cellular base stations. Specifically, DeepAR achieved lower \ac{MSE}, \ac{MASE}, \ac{MAPE}, \ac{QL}, \ac{ND} and higher Coverage values, indicating its ability to provide more accurate and reliable estimates of \ac{PRB} utilization. At the end of the paper, we also highlighted some important aspects of using probabilistic forecast methods in the context of deep learning models such as DeepAR, including understanding the limitations associated with small data sets, training time, complexity, and suitability for non-stationary data sets.


\bibliographystyle{ieeetr}
\bibliography{biblio}  

\balance

\end{document}